\documentclass[11pt,twoside]{article}
\usepackage[table]{xcolor}
\usepackage{amsmath,amsthm,amssymb,amsbsy,epsfig,fancyhdr,psfrag,calc,ifthen,float,mathrsfs,hhline}
\PassOptionsToPackage{ctagsplt,Righttag}{amstex}
\usepackage[colorlinks=true,citecolor=blue,linkcolor=blue,
            bookmarksnumbered=true,pdfstartview={FitH},
            hyperindex=true,linktoc=all,linktocpage=true]{hyperref}
\usepackage{mathtools}
\usepackage{framed}
\usepackage{dsfont}
\usepackage{ulem}
\usepackage{tabularx}
\usepackage{booktabs}
\usepackage{amsmath}
\usepackage{tikz}
\usepackage{enumitem}
\usepackage{caption}
\usepackage{epsfig}
\usepackage{cite}

\newcommand{\bC}{\mbox{\boldmath $C$}}

\newcommand{\bI}{\mbox{\boldmath $I$}}

\newcommand{\bd}{\mbox{\boldmath $d$}}
\newcommand{\be}{\mbox{\boldmath $e$}}

\newcommand{\bn}{\mbox{\boldmath $n$}}

\newcommand{\bp}{\mbox{\boldmath $p$}}

\newcommand{\bs}{\mbox{\boldmath $s$}}

\newcommand{\Bepsilon}{\mbox{\boldmath $\epsilon$}}

\newcommand{\Beta}{\mbox{\boldmath $\eta$}}

\newcommand{\Bxi}{\mbox{\boldmath $\xi$}}

\newcommand{\Bsigma}{\mbox{\boldmath $\sigma$}}

\newcommand{\Bphi}{\mbox{\boldmath $\phi$}}

\setlist[itemize]{noitemsep,nolistsep} 
\usepackage[boxed]{algorithm2e}
\usepackage[flushleft]{threeparttable}
 
\SetAlCapSkip{1em}
\usepackage{color}
\definecolor{darkgreen}{rgb}{0,0.65,0}
\definecolor{darkred}{rgb}{0.65,0,0}

\newcommand{\df}[2]{\displaystyle\frac{d #1}{d #2}}
\newcommand{\partialf}[2]{\displaystyle\frac{\partial #1}{\partial #2}}
\newcommand{\partialPP}[2]{\displaystyle\frac{\partial^2 #1^{#2}}{\partial p_i \partial p_j}}
\newcommand{\partialPE}[2]{\displaystyle\frac{\partial^2 #1^n}{\partial p_i \partial \xi_k^{#2}}}
\newcommand{\partialEP}[2]{\displaystyle\frac{\partial^2 #1^n}{\partial \xi_k^{#2} \partial p_j}}
\newcommand{\partialEE}[3]{\displaystyle\frac{\partial^2 #1^n}{\partial \xi_k^{#2} \partial \xi_m^{#3}}}
\newcommand{\dPP}[1]{\displaystyle\frac{d^2 #1^n}{d p_i d p_j}}
\newcommand{\dEP}[3]{\displaystyle\frac{d \xi_{#1}^{#3}}{d p_{#2}}}
\def\lsp{\def\baselinestretch{0.75}\large\normalsize}
\def\ssp{\def\baselinestretch{1.0}\large\normalsize}
\def\dsp{\def\baselinestretch{1.37}\large\normalsize}
\advance\textheight by \topskip
\begin{document}
\bibliographystyle{ieeetr}
\flushbottom
\pagestyle{empty}
\pagenumbering{arabic}
\ssp
\title{\vspace{-0.75in}\bf A Direct-adjoint Approach for Material Point Model Calibration 
with Application to Plasticity} 
\vspace{0.1in}
\author{\hspace{-0.25in} Ryan YAN$^{1}$,\ D. Thomas SEIDL$^{2,}$\footnote{dtseidl@sandia.gov (corresponding author)},\ Reese E. JONES$^{3}$,\ and Panayiotis PAPADOPOULOS$^{1}$
\\[0.05in]
\sl\small $^1$Department of Mechanical Engineering, University of California,
Berkeley, CA 94720-1740, USA \\
\sl\small $^2$Sandia National Laboratories, Albuquerque, NM 87123, USA\\
\sl\small $^3$Sandia National Laboratories, Livermore, CA 94550, USA\\
\rm \normalsize}
\date{}
\maketitle
\small
\lsp
\par\noindent
\protect\vspace{-0.6in}
\renewcommand{\contentsname}{\normalsize\centerline{Table of contents}}
\tableofcontents
\ssp
\protect\vspace{0.025in}
\normalsize
\begin{abstract}
This paper proposes a new approach for the calibration of material parameters in local elastoplastic constitutive models. The calibration is posed as a constrained optimization problem, where the constitutive model evolution equations for a single material point serve as constraints. The objective function quantifies the mismatch between the stress predicted by the model and corresponding experimental measurements. 
To improve calibration efficiency, a novel direct-adjoint approach is presented to compute the Hessian of the objective function, which enables the use of second-order optimization algorithms. Automatic differentiation is used for gradient and Hessian computations. Two numerical examples are employed to validate the Hessian matrices and to demonstrate that the Newton-Raphson algorithm consistently outperforms gradient-based algorithms such as L-BFGS-B. 
\end{abstract}
\vspace{0.075in}
\noindent{\textbf{Keywords:}
adjoint method; automatic differentiation; model calibration; plasticity} 
\newpage
\dsp
\pagestyle{fancyplain}
\section{Introduction}\label{sec:intro}
\par\noindent

Constitutive models, also commonly known as material models, play an essential role in computational simulations of structural or mechanical systems. Every constitutive model has a number of free parameters that must be determined using experimental data from mechanical \emph{characterization} experiments through a process known as \emph{calibration} \cite{ma16020836}. Improvements in calibration lead to more accurate simulations and reliable predictions. While experimental methods for mechanical characterization have become quite advanced, robust calibration techniques are desperately lacking and represent an impediment in the use of sophisticated constitutive models.
\par
Most calibration problems are posed through an optimization-based formulation,
where the goal is to minimize an objective function that measures the
discrepancy between simulation predictions and experimental measurements
by finding an optimal set of constitutive model parameters~\cite{article}. Various techniques
for performing this optimization exist and can be grouped by
the amount of derivative information they use in the optimization step.
Some studies resort to using gradient-free methods including genetic algorithms \cite{ROBSON2024104881, PAL1996325, ROKONUZZAMAN2010573, Levasseur, papon}, particle-swarm algorithms \cite{doi:10.1061/(ASCE)EM.1943-7889.0001214}, or Rosenbrock methods \cite{https://doi.org/10.1002/nag.137}. Methods that only use values of the objective function are simple to implement, but could require thousands of objective function evaluations to reach convergence. Consequently, for large-scale calibration problems involving nonlinear finite element (FE) solves, relying solely on gradient-free algorithms can result in significant computational cost. Other studies have calibrated constitutive models by employing quasi-Newton gradient-based methods such as BFGS \cite{CEKEREVAC2006432, app9091799}. Gradient-based methods are generally considered more effective than those relying solely on objective function values. However, the process of computing the gradient introduces its own set of challenges. One widely used approach to differentiating the objective function with respect to the model parameters is the forward finite difference method \cite{SAKARIDIS2024113076, Yueqi, VARGAS20226733, Cooreman}. A significant drawback of using finite differences is that calculating each gradient component necessitates a complete FE solution. To address this limitation, a computationally efficient method for obtaining the gradient of the calibration objective function using adjoint sensitivities is presented in \cite{seidl_calibration_2022}. 
\par
Generally, optimization methods that utilize higher-order derivatives tend to converge in fewer iterations~\cite{numericalopt}. To date, the use of \emph{Hessian} (second-order derivatives) information for material model calibration is lacking, largely because finite-difference approximations of the Hessian
are prohibitively expensive to obtain. In an attempt to circumvent this problem, some studies make the assumption that near the optimum, the Hessian can be approximated by discarding terms involving second-order derivatives \cite{SAKARIDIS2024113076, NEGGERS201971}. In \cite{ARIAN1999853, SHIDONG2018265}, the adjoint method is used to approximate the Hessian in the vicinity of the optimal solution. However, to achieve true quadratic convergence in an optimization algorithm such as the Newton-Raphson method, the exact Hessian is required. In view of the above, this study aims to advance calibration through efficient computation of the exact parameter Hessian, the matrix of second-order derivatives of the objective function.        
\par
Practically, the calibration problem is solved numerically through the introduction of computational model of the characterization experiment. FE models, which are derived from both conservation laws and constitutive models, are commonly utilized for this purpose, but in some scenarios a material point model is sufficient. Material point models are local statements of the constitutive model that can be also interpreted as complete descriptions of a body undergoing homogeneous deformation under suitable external loading. Uniaxial tension of a materially homogeneous bar, where all points in the body experience the same stress state, is the prototypical example of a material point model, but there exist other useful such states, such as simple shear and dilation. We recommend Reference \cite{lame_manual} for additional examples. The local state variables present in a material point model are constitutive model-specific, although there are often multiple ways to define them for a given constitutive model \cite{alberdi_unified_2018}. For elastic-plastic models with isotropic hardening, the isotropic hardening variable or equivalent plastic strain (for small-strain models) is an example of a local state variable.

\par
This study lays the foundation for future development of a method to compute the parameter Hessian
for the calibration problem with a FE model constraint by examining the simpler material point problem, where the constraint simplifies to the local residual at a single quadrature point. 
In this context, a direct-adjoint approach to computing the Hessian is presented by following the approach taken in~\cite{papadimitriou}, where they did so for a steady-state PDE. Numerous checks of accuracy for the direct-adjoint Hessian are provided. Upon verification, the Hessian is applied to solve synthetic and experimental calibration problems, which shows its usefulness even in this limited context. Of course, the transition of calibration for a local material point model to a global FE representation of a boundary value problem will require comparison to experimentally observable quantities (e.g.\ full-field displacements and load cell measurements), which will affect the objective function and increase the complexity of the derivation of the associated Hessian.
\par
The remainder of the paper is structured as follows: Section~\ref{sec:Forward_Model} contains an introduction to
the exemplar constitutive model under consideration in this work, a small deformation elastic-plastic model.
In Section~\ref{sec::Inverse Problem Formulation}, the direct and adjoint methods for computing the gradient are reviewed. These
are then utilized in the novel direct-adjoint formulation for the Hessian matrix. Section~\ref{sec:numerical_results} documents results from the application of our approach to synthetic plane stress
and experimental uniaxial tension examples. Finally, Section~\ref{sec:con} summarizes and
concludes.
\section{Material Point Plasticity Forward Model}\label{sec:Forward_Model}
\par\noindent
This section describes the forward model of plasticity that will be calibrated in the ensuing developments. Here, the constitutive equations are defined and then compiled into a set of residual equations that comprise the overall plasticity model. Finally, 
the local state variables are computed at each discrete load step to satisfy the residual equations.
\subsection{Constitutive equations}
\par\noindent
In this study, an infinitesimal-deformation plasticity model is considered. 
To this end, the total strain~$\Bepsilon$ is assumed to be additively decomposed as 
\begin{equation}
    \Bepsilon\ =\ \Bepsilon^{e} + \Bepsilon^{p}\ ,
\end{equation}
where~$\Bepsilon^{e}$ and~$\Bepsilon^{p}$ denote the elastic and plastic strains, respectively. Below the yield limit, the material response is assumed to be isotropic and linearly elastic. Thus, the stress~$\Bsigma$ is written as
\begin{equation} \label{eq:elastic_stress}
\Bsigma\ =\ 
\lambda \, \text{tr}\left[\Bepsilon^e\right]\bI + 2 \mu \Bepsilon^e\ .
\end{equation}
Here, $\lambda = \frac{E\nu}{(1+\nu)(1- 2\nu)}$ and~$\mu=\frac{E}{2(1+\nu)}$ 
are the Lam\'e constants, expressed in terms of the Young's modulus, $E$, and  
the Poisson's ratio, $\nu$, while $\bI$ denotes the second-order identity tensor. 
\par
A yield function $f$ is introduced in the form
\begin{equation}
    f(\Bsigma, \alpha)\ =\ \phi(\Bsigma) - \bar{\sigma}(\alpha)\ ,
\end{equation}
where~$\phi(\Bsigma)$ is the effective stress and $\alpha$ is the equivalent plastic strain. 
The current yield stress $\bar{\sigma}(\alpha) = Y + H(\alpha)$ is the sum of the initial yield limit~$Y$, a linear hardening term, and a Voce-type hardening term, the latter two jointly expressed as  
\begin{equation}\label{eq:H}
    H(\alpha)\ =\ K\alpha + S(1 - \text{exp}\bigl(-D\alpha)\bigr)\ .
\end{equation}
The material constants~$K$, $S$, and~$D$ in Equation~\eqref{eq:H} represent the linear hardening modulus, the hardening saturation modulus, and the hardening saturation rate,
respectively. In this work, a von Mises-type effective stress~$\phi$ is adopted, such that 
\begin{equation} \label{eq:effective_stress}
    \phi(\Bsigma)\ =\ \sqrt{\frac{3}{2}}\lVert \bs \lVert\ ,
\end{equation}
where $\bs = \Bsigma - \frac{\text{tr}[\Bsigma]}{3}\bI$ is the deviatoric stress and 
$\lVert \bs \lVert = (\bs:\bs)^{1/2}$. The stress and the equivalent plastic strain satisfy the condition~$f(\Bsigma,\alpha) \leq 0$ at all times, where~$f < 0$ implies an elastic material response (loading or unloading), while~$f = 0$ implies plastic or neutral loading. 
A strain-space representation of the yield function will be necessary later and can be effected with the aid of Equation~\eqref{eq:elastic_stress}, by defining a new function~$g$ such that
\begin{equation} \label{eq:yield_func_strain}
    f(\Bsigma,\alpha) \ =\ g(\Bepsilon - \Bepsilon^{p}, \alpha) \ .
\end{equation}
\par
An associative flow rule for the evolution of the plastic strain is defined as 
\begin{align} \label{eq:flow_rule}
    \dot{\Bepsilon}^p &\ =\ \gamma\frac{\partial f}{\partial\bs}\ =\ \gamma \bn\ ,
\end{align}
so that $\dot{\Bepsilon}^p$ is always normal to the yield surface in deviatoric stress space \cite{computationalplasticity}. The plastic multiplier~$\gamma$ is subject to the Kuhn-Tucker loading/unloading conditions 
\begin{align} 
    \gamma\ \geq\ 0\quad ,\quad f\ \leq\ 0\quad ,\quad \gamma f\ =\ 0\ .
\end{align}
In view of the form of the effective stress in Equation~\eqref{eq:effective_stress}, 
it follows from Equation~\eqref{eq:flow_rule} that $\text{tr}[\dot{\Bepsilon}^p] = 0$, 
hence also that 
\begin{equation} \label{eq:plastic_incompressibility}
    \text{tr}[\Bepsilon^p]\ =\ 0\ .
\end{equation}
This means that the plastic deformation is volume-preserving.
Lastly, the evolution of the equivalent plastic strain is governed by the rule
\begin{equation}\label{eq:alpha-dot}
    \dot{\alpha}\ =\ \gamma\ .
\end{equation}
\par
The constitutive model comprises six independent material parameters: $E$, 
$\nu$, $K$, $Y$, $S$, and~$D$. These constants, written as components of a vector $\bp$, serve as the optimization parameters in the calibration of the constitutive model. 
\subsection{Model structure} \label{sec:model_structure}
\par\noindent
The material point model of elastoplasticity is cast here as a set of residual equations in terms of the local state variables and constitutive model parameters. Upon formulation of the constitutive equations, the model itself takes as
input the strain~$\Bepsilon$ at a discrete number of load steps $n = 1, \ldots, N_L$. In this regard, the model can be thought of as being strain-driven with the stress being determined at every load step as a function of strain. At each 
step~$n$, a vector $\Bxi^{n}$ of local state variables is defined, which contains the components of the plastic strain tensor~$\Bepsilon^p$ and the equivalent plastic strain~$\alpha$, in the form
\begin{equation}
  [\epsilon^p_{11},\ \epsilon^p_{12},\ \epsilon^p_{13},\
\epsilon^p_{22},\ \epsilon^p_{23},\ \epsilon^p_{33},\ \alpha]^{n}, \quad n\ =\
1,\dots,N_L\ ,
\end{equation}
where~$N_L$ denotes the total number of load steps. Additionally, note here 
that $\epsilon^p_{33} = -\left(\epsilon^p_{11} + 
\epsilon^p_{22}\right)$ due to the plastic incompressibility 
condition~\eqref{eq:plastic_incompressibility}, so this variable is not included in the vector~$\Bxi^n$. 
\par
To ensure that the state variables satisfy the constitutive equations, the model requires a set~$\bC^n$ of local residual functions that provides a complete mathematical description of the constitutive model. Taking into account Equations~\eqref{eq:flow_rule} and \eqref{eq:alpha-dot}, this set takes the form 
\begin{equation} \label{eq:local_residual}
\bC^n(\Bxi^{n}, \Bxi^{n-1}, \bp)\ = \ \left\{
\begin{aligned}
&f\ <\ 0: \left\{
\begin{aligned}
  &\left(\Bepsilon^p\right)^n - \left(\Bepsilon^p\right)^{n-1}\ , \\
  &\alpha^n - \alpha^{n-1}\ .
\end{aligned} \right. \\
&f\ =\ 0: \left\{
\begin{aligned}
  &\left(\Bepsilon^p\right)^n - \left(\Bepsilon^p\right)^{n-1}
  - \left(\alpha^n - \alpha^{n-1}\right)\bn^n\ , \\
  &g(\Bepsilon^{n} - (\Bepsilon^{p})^n, \alpha^n)\ ,
\end{aligned} \right.
\end{aligned}
\right.
\end{equation}
where the function $g$ is defined in Equation~\eqref{eq:yield_func_strain}. 
In~\eqref{eq:local_residual}, a backward Euler integration method is employed for the rate-type flow equations during plastic loading~\cite{SimoHughes}.
It follows from Equation~\eqref{eq:local_residual} that, for given~$\Bxi^{n-1}$
and~$\bp$, $\Bxi^{n}$ must satisfy 
\begin{equation} \label{eq:residual_eqn}
    \bC^n(\Bxi^{n}, \Bxi^{n-1}, \bp)\ =\ \mathbf{0}, \quad n\ =\ 1,\dots,N_L\ .
\end{equation}
\par
The Newton-Raphson method is used to solve 
Equation~\eqref{eq:residual_eqn} for~$\Bxi^n$ at each load step~$n \in [1, N_L]$. 
Each iteration of the algorithm corresponds to the linear system 
\begin{equation} \label{eq::newton_solve}
    \frac{\partial\bC^n}{\partial\Bxi^n}\Delta\Bxi^n\ =\ -\bC^n(\Bxi^{n},
\Bxi^{n-1}, \bp)\ , \quad n\ =\ 1,\dots,N_L\ ,
\end{equation}
where $\Bxi^{n+1} = \Bxi^{n} + \Delta\Bxi^{n}$. Equation~\eqref{eq::newton_solve} is subject to the initial condition $\Bxi^0 = \mathbf{0}$, reflecting the state of no plastic deformation. After determining the local state variables, the stress tensors~$\Bsigma^n$ 
needed to evaluate the objective function are computed using 
Equation~\eqref{eq:elastic_stress}. 
\section{Inverse Problem Formulation}\label{sec::Inverse Problem Formulation}
\par\noindent
In this section, the objective function is introduced for the material point model calibration problem. This function is defined in terms of differences of the model stress from the experimentally (or synthetically) generated one at different loading states. The \textit{direct} and \textit{adjoint} methods for computing the gradient of the objective function are presented and discussed. In addition to obtaining the gradient, a primary contribution of this work is to efficiently compute the Hessian of the objective function through the so-called \textit{direct-adjoint} approach 
\cite{papadimitriou}. The JAX automatic differentiation (AD) library~\cite{jax} is used to obtain partial derivatives of the residual and objective functions with respect to the local state variables and material parameters. 
\subsection{Constrained optimization formulation}
\par\noindent
The goal of material model calibration is to minimize an objective function~$J$ that 
quantifies the mismatch between the stress~$\Bsigma$ with corresponding 
experimental measurements~$\tilde{\Bsigma}$ at each load step. This function is defined here as 
\begin{equation}
    J(\Bxi,\bp)\ =\ \sum_{n = 1}^{N_L}J^n(\Bxi^n,\bp)\
=\ \sum_{n = 1}^{N_L}\frac{1}{2}\lVert\Bsigma^n(\Bxi^n,\bp) -
\tilde{\Bsigma}^n\rVert^2\ ,
\end{equation}
so that the optimization problem may be expressed as 
\begin{equation*}
    \min_{\bp} \quad J(\Bxi,\bp)\ ,
\end{equation*}
\begin{equation*}
    \quad \bC^n(\Bxi^{n}, \Bxi^{n-1}, \bp)\ =\ \mathbf{0}, \quad n\
=\ 1,\dots,N_L\ ,
\end{equation*}
\begin{equation}\label{eq:opt_eqn}
    \bp_{\min}\ \leq\ \bp\ \leq\ \bp_{\max}\ ,
\end{equation}
where $\bp_{\min}$ and $\bp_{\max}$ denote the lower and upper optimization search bounds for the material parameters in~$\bp$. The constraints in Equation~\eqref{eq:opt_eqn} must be satisfied at each iteration 
of the optimization algorithm, which amounts to one complete solution of the forward 
problem per iteration. For minimization, the L-BFGS-B optimization algorithm, as implemented by the Scientific Python (SciPy) library \cite{scipy}, is initially used. This quasi-Newton algorithm requires evaluations of the objective function and its gradient. In this work, \textit{direct} and \textit{adjoint} local sensitivity analyses are employed to compute the gradient of the objective function. The convergence of the L-BFGS-B method is compared in Section~\ref{sec:numerical_results} to that of the Newton-Raphson algorithm, which in addition to the gradient, requires the Hessian of the objective function.
\par
The procedure for solving Equation~\eqref{eq:opt_eqn} begins with an initial guess, denoted as~$\bp^{(0)}$. Given this fixed value of~$\bp$, the gradient and Hessian of the objective function are computed using one of the methods described next. A gradient or Hessian based optimization algorithm such as L-BFGS-B or Newton-Raphson is employed to update~$\bp$. This iterative process continues until the solution converges to a minimum.    
\subsection{Direct method}
\par\noindent
The first approach employed in this work to obtain the objective function gradient is referred to as the \textit{direct} method. Treating the local state variables~$\Bxi$ as 
implicit functions of~$\bp$, the objective function and residual equations are 
differentiated with respect to~$\bp$ to obtain 
\begin{equation}\label{eq:diff_J}
    \df{J}{p_i}\ =\ \sum_{n = 1}^{N_L}
\left(\partialf{J^n}{\xi_k^n}\df{\xi_k^n}{p_i} + \partialf{J^n}{p_i}\right)\ ,
\end{equation}
\begin{equation}\label{eq:diff_C}
    \df{C_q^n}{p_i}\ =\ \partialf{C_q^n}{\xi_k^n}\df{\xi_k^n}{p_i} +
\partialf{C_q^n}{\xi_k^{n-1}}\df{\xi_k^{n-1}}{p_i} + \partialf{C_q^n}{p_i}\ =\
0\ .
\end{equation}
Note that in Equations~(\ref{eq:diff_J}, \ref{eq:diff_C}) and all subsequent mathematical
expressions, the index ranges are defined as follows: $q \in [1, N]$, $i,j \in [1, 
P]$, and $k,m \in [1, M]$, where~$N$ is the number of residual equations, $P$ 
is the number of material parameters, and~$M$ is the number of local state 
variables. It is important to note here that $M = N$, which reflects a necessary condition for the solution of the residual equations. Moreover, load steps denoted by the superscript $n$ are explicitly summed, while all other indices follow the Einstein summation convention. Finally, the matrix~$\left[\df{\xi_k^n}{p_i}\right]$ is computed at each load step 
by solving the system of linear equations obtained from Equation~\eqref{eq:diff_C}, 
leading to 
\begin{equation}\label{eq:fwd_sens_mat}
    \df{\xi_k^n}{p_i}\ =\ -\left(\partialf{C_q^n}{\xi_k^n}\right)^{-1}\left(\partialf{C_q^n}{\xi_k^{n-1}}\df{\xi_k^{n-1}}{p_i} + \partialf{C_q^n}{p_i}\right), \quad n\ =\ 1,
\ldots,N_L\ .
\end{equation}
Execution of the direct approach for computing the gradient may be combined with that of the forward solution. Additionally, this method only requires the storage of state variables and corresponding matrices from two adjacent load steps. However, the computational cost of the direct approach increases with the dimension of the material parameter space. More specifically, $P$ systems of linear algebraic equations are required to solve Equation~\eqref{eq:fwd_sens_mat} at each load step. 
For this reason, the so-called \textit{adjoint} approach is considered as an alternative. Although the adjoint method requires extra memory, its computational cost is independent of the dimension of the material parameter space.
\subsection{Adjoint method}
\par\noindent
To derive the gradient of the objective function using the adjoint method, a new 
functional~$\hat{J}$ is defined as 
\begin{equation}\label{eq:lagrangian}
    \hat{J}(\Bxi^n, \bp, \Bphi^n)\ =\ \sum_{n=1}^{N_L}\left(J^n(\Bxi^n,\bp) + (\Bphi^n)^T
\bC^n(\Bxi^{n}, \Bxi^{n-1}, \bp)\right)\ ,
\end{equation}
where $\Bphi^n$ is the vector of the \textit{adjoint local state variables}.
Differentiating~$\hat{J}$ with respect to~$\bp$ yields 
\begin{equation}\label{eq:diff_J_hat}
    \df{\hat{J}}{p_i}\ =\ \sum_{n = 1}^{N_L}\left(\df{J^n}{p_i} + \phi^n_q
\df{C_q^n}{p_i}\right)\ ,
\end{equation}
where~$\df{J^n}{p_i}$ and~$\df{C_q^n}{p_i}$ are given by 
Equations~(\ref{eq:diff_J},\ref{eq:diff_C}). Substituting these into 
Equation~(\ref{eq:diff_J_hat}) and rearranging terms results in 
\begin{equation}\label{eq:Lagrangian_derivative}
    \df{\hat{J}}{p_i}\ =\ \sum_{n = 1}^{N_L}\left[\partialf{J^n}{p_i} +
\phi^n_q \partialf{C_q^n}{p_i} + \left(\phi^n_q \partialf{C_q^n}{\xi_k^n} +
\partialf{J^n}{\xi_k^n}\right)\df{\xi_k^n}{p_i} + \phi^n_q
\partialf{C_q^n}{\xi_k^{n-1}}\df{\xi_k^{n-1}}{p_i}\right]\ .
\end{equation}
Upon expanding the summation in Equation~\eqref{eq:Lagrangian_derivative}, one 
notices that computing explicitly all~$\df{\xi_k^n}{p_i}$ terms can be avoided if 
\begin{equation}\label{eq:adj_eqn_NL}
    \partialf{J^{N_L}}{\xi_k^{N_L}} +
\phi_q^{N_L}\partialf{C_q^{N_L}}{\xi_k^{N_L}}\ =\ 0
\end{equation}
and
\begin{equation}\label{eq:adj_eqns}
    \partialf{J^n}{\xi_k^n} + \phi_q^n\partialf{C_q^n}{\xi_k^{n}} +
\phi_q^{n+1}\partialf{C_q^{n+1}}{\xi_k^{n}}\ =\ 0, \quad n\ =\ N_L - 1,\dots,1\ .
\end{equation}
In direct notation, Equations~(\ref{eq:adj_eqn_NL},\ref{eq:adj_eqns}) take the 
form 
\begin{equation}\label{eq:adj_eqn_NL_direct}
    \underbrace{\left(\partialf{J^{N_L}}{\Bxi^{N_L}}\right)^T}_{(M \times 1)} +
\underbrace{\left(\partialf{\bC^{N_L}}{\Bxi^{{N_L}}}\right)^T}_{(M \times N)}
\underbrace{\Bphi^{N_L}}_{(N \times 1)}\ =\ \mathbf{0}
\end{equation}
and
\begin{equation}\label{eq:adj_eqns_direct}
    \underbrace{\left(\partialf{J^n}{\Bxi^n}\right)^T}_{(M \times 1)} + \underbrace{\left(\partialf{\bC^n}{\Bxi^{n}}\right)^T}_{(M \times N)} 
    \underbrace{\Bphi^n}_{(N \times 1)} +
\underbrace{\left(\partialf{\bC^{n+1}}{\Bxi^{n}}\right)^T}_{(M \times N)}
\underbrace{\Bphi^{n+1}}_{(N \times 1)}\ =\ \mathbf{0}, \quad n\ =\ N_L -
1,\dots,1\ ,
\end{equation}
respectively. The above are known as the \textit{adjoint equations}, and are solved 
backwards in time for $\Bphi^n$, beginning at load step~$N_L$ and ending at step 1. 
Therefore, the forward problem must be solved first so that the entire history of the state variables be known. After solving for the adjoint variables at each load step, the gradient now takes the simplified form 
\begin{equation}
    \df{J}{p_i}\ =\ \sum_{n = 1}^{N_L}\left(\partialf{J^n}{p_i} + \phi^n_q
\partialf{C_q^n}{p_i}\right)\ .
\end{equation}
\par
The computational complexity of the adjoint method is independent of the dimension of the material parameter space because only one system solution is required to compute $\Bphi^n$ at each step. However, to perform the reverse recursion, one must allocate additional memory to store the local state variables over all the load steps.     
\subsection{Direct-adjoint method}
\par\noindent
As argued in \cite{papadimitriou}, the most computationally efficient 
sensitivity-based scheme for obtaining the objective function Hessian is 
referred to as the \textit{direct-adjoint} approach. First, the functional 
$\hat{J}$ of Equation~\eqref{eq:lagrangian} is twice-differentiated with respect to~$\bp$, leading to
\begin{equation}\label{eq:diff2_J_hat}
    \frac{d \hat{J}}{d p_i d p_j}\ =\ \sum_{n = 1}^{N_L} \left(\dPP{J} +
\phi^n_q \dPP{C_q}\right)\ .
\end{equation}
The terms~$\dPP{J}$ and~$\dPP{C_q}$ are obtained by differentiating 
Equations~(\ref{eq:diff_J},\ref{eq:diff_C}) with respect to~$\bp$, resulting in 
\begin{equation}
\begin{aligned}[b]
    \frac{d^2J}{dp_i dp_j} \ &=\ \sum_{n = 1}^{N_L} \left(\partialPP{J}{n} +
\partialPE{J}{n} \dEP{k}{j}{n} + \partialEP{J}{n}\dEP{k}{i}{n} +
\partialEE{J}{n}{n}\dEP{k}{i}{n}\dEP{m}{j}{n}+
\partialf{J^n}{\xi_k^n}\partialPP{\xi_k}{n}\right)
\end{aligned}
\label{eq:diff2_J}
\end{equation}
and
\begin{equation}
\begin{aligned}[b]
    \dPP{C_q} \ =&\ \partialPP{C_q}{n} + \partialPE{C_q}{n}\dEP{k}{j}{n} + \partialPE{C_q}{n-1}\dEP{k}{j}{n-1}+\partialEE{C_q}{n}{n}\dEP{k}{i}{n}\dEP{m}{j}{n}\\[0.2cm]&+\partialEE{C_q}{n}{n-1}\dEP{k}{i}{n}\dEP{m}{j}{n-1}+ \partialEP{C_q}{n}\dEP{k}{i}{n} + \partialf{C_q^n}{\xi_k^n}\partialPP{\xi_k}{n} + \partialEE{C_q}{n-1}{n}\dEP{k}{i}{n-1}\dEP{m}{j}{n} \\[0.2cm]&+ \partialEE{C_q}{n-1}{n-1}\dEP{k}{i}{n-1}\dEP{m}{j}{n-1}
    + \partialEP{C_q}{n-1}\dEP{k}{i}{n-1} +
\partialf{C_q^n}{\xi_k^{n-1}}\partialPP{\xi_k}{n-1}\ .
\end{aligned}
\label{eq:diff2_C}
\end{equation}
Substituting Equations~(\ref{eq:diff2_J},\ref{eq:diff2_C}) into 
Equation~\eqref{eq:diff2_J_hat}, one can again observe that explicit 
computation of all~$\partialPP{\xi_k}{n}$ terms can be avoided if 
Equations~(\ref{eq:adj_eqn_NL_direct}, \ref{eq:adj_eqns_direct}) are satisfied. 
After solving for each~$\Bphi^n$, the Hessian matrix components are given by 
\begin{equation}
\begin{aligned}[b]
\frac{d^2 J}{d p_i d p_j} \ =&\ \sum_{n = 1}^{N_L} \Biggl[\partialPP{J}{n} +
\phi_q^n\partialPP{C_q}{n} + \left(\partialPE{J}{n} +
\phi_q^n\partialPE{C_q}{n}\right)\dEP{k}{j}{n} + \left(\partialEP{J}{n} +
\phi_q^n\partialEP{C_q}{n}\right)\dEP{k}{i}{n}\\[0.2cm] &+
\left(\partialEE{J}{n}{n} +
\phi_q^n\partialEE{C_q}{n}{n}\right)\dEP{k}{i}{n}\dEP{m}{j}{n} +
\phi_q^n\left(\partialPE{C_q}{n-1}\dEP{k}{j}{n-1} +
\partialEE{C_q}{n}{n-1}\dEP{k}{i}{n}\dEP{m}{j}{n-1} \right.\\[0.2cm] &\left.+
\partialEE{C_q}{n-1}{n}\dEP{k}{i}{n-1}\dEP{m}{j}{n} +
\partialEE{C_q}{n-1}{n-1}\dEP{k}{i}{n-1}\dEP{m}{j}{n-1} +
\partialEP{C_q}{n-1}\dEP{k}{i}{n-1}\right)\Biggr]\ ,
\end{aligned}
\end{equation}
where ~$\dEP{k}{i}{n}$ terms are computed using 
Equation~\eqref{eq:fwd_sens_mat}, and all second derivatives are computed efficiently 
using~AD. For each load step, the computational cost 
is~$P$ systems of linear equations to compute~$\dEP{k}{i}{n}$, 
and an additional one to solve the adjoint equations. Over all load steps, 
$N_L(P + 1)$ linear system solutions are required to compute the objective 
function Hessian matrix. 
\par
As stated earlier, the Newton-Raphson method requires 
both the objective function gradient and Hessian. It follows that one 
attractive property of the direct-adjoint method is that the adjoint 
equations are identical to those in the adjoint approach for the gradient. 
Therefore, the adjoint variables~$\Bphi^n$ obtained while solving for the 
gradient can be reused in computing the Hessian. 
\subsection{Parameter scaling}
\par\noindent
Before running an optimization algorithm such as L-BFGS-B, it is generally recommended to scale the model's material parameters \cite{numericalopt}. The scaled parameter vector, denoted~$\Beta$, is related to the original parameter vector~$\bp$ by the natural-logarithmic relation
\begin{equation} \label{eq:parameter_scaling}
    \eta_i\ =\ \ln \left(\frac{p_i}{p_{i,\text{ref}}}\right)\ ,
\end{equation}
where~$p_{i,\text{ref}}$ denotes the reference parameter value of the $i$-th element of~$\bp$. In this case, $p_{i,\text{ref}}$ is chosen to be the initial guess for the parameter~$p_i$.
\par
To effectively use the scaled parameters in the optimization algorithms, the 
gradient and Hessian must also be transformed into~$\Beta$-space. Employing the 
chain rule, the transformed components of the gradient are resolved as 
\begin{equation} \label{eq:transformed_grad}
    \frac{dJ}{d\eta_i}\ =\ \frac{dJ}{dp_r}\frac{dp_r}{d\eta_i}\ .
\end{equation}
Differentiating Equation~\eqref{eq:transformed_grad} with respect to~$\eta$, the transformed Hessian components are given by 
\begin{equation}
    \frac{d^2J}{d\eta_id\eta_j}\ =\
\frac{d^2J}{dp_rdp_s}\frac{dp_r}{d\eta_i}\frac{dp_s}{d\eta_j} +
\frac{dJ}{dp_r}\frac{d^2p_r}{d\eta_id\eta_j}\ ,
\end{equation}
where~$r,s \in [1, P]$. 
%
\section{Numerical Results}\label{sec:numerical_results}
\par\noindent
In this section, the proposed sensitivity-based approaches are applied to two different 
problems. The first problem utilizes synthetically generated data to provide 
verification of the AD-computed gradients and Hessian matrices. The second 
problem demonstrates model calibration using experimental tension 
test data. Both problems highlight the effectiveness of second-order 
Hessian-based optimization algorithms for inverse problems, in contrast to 
methods such as L-BFGS-B that rely solely on the gradient. Comparisons between the Newton-Raphson method and L-BFGS-B are made in terms of the number of iterations required for convergence. The total computational cost is estimated based on the required number of system solutions. 
\subsection{A plane-stress problem}
\par\noindent
To demonstrate the validity of the objective function gradients and Hessians 
obtained using direct and adjoint approaches, a synthetic plane-stress problem 
with Voce-type hardening is considered. In this problem, a material 
particle is loaded in biaxial stress on a plane with normal vector~$\be_3$, 
where vectors \{$\be_1$,~$\be_2$,~$\be_3$\} form an orthonormal basis. During 
the first 50 load steps, the material is fixed in the~$\be_2$-direction while 
being stretched to a maximum strain of 0.02 in the~$\be_1$-direction. In the 
subsequent 50 load steps, the material is fixed at 0.02 strain in the 
$\be_1$-direction and stretched to 0.02 strain in the~$\be_2$-direction. To 
satisfy the plane-stress conditions, several adjustments were made to the 
model structure from Section \ref{sec:model_structure}. These modifications are 
detailed in Appendix~\ref{App:plane}. 
\par
The actual material model parameters $\bp$ are chosen to be 
\begin{equation}
    [E, \nu, Y, K, S, D]\ =\ [70000\text{, } 0.3\text{, }
200\text{, } 0\text{, } 200\text{, } 20]\ ,
\end{equation}
where $E$, $Y$, $K$, and $S$ are presented in units of MPa. For this example, the elastic material parameters~$E$, $\nu$ and the linear hardening modulus~$K$ are assumed to be known, which means that they are fixed in value and are not considered as design variables for the optimization problem. 
\par
\begin{figure}[H]
    \centerline{\includegraphics[scale= 0.60]{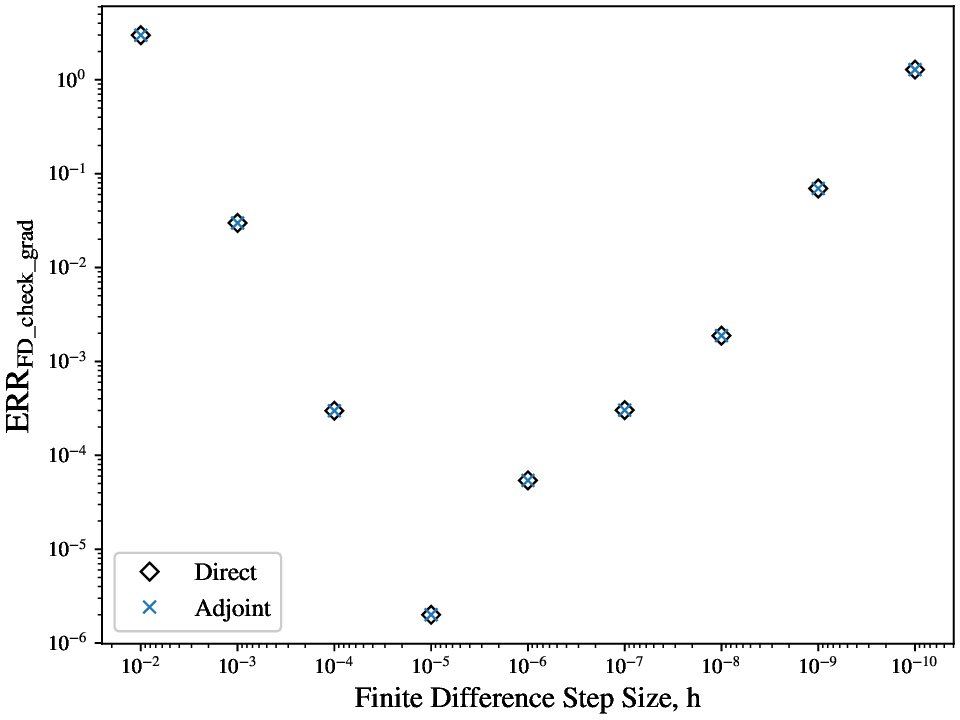}}
    \caption{Gradient comparisons to finite difference gradients}
    \label{fig:FD_grad}
\end{figure}
\begin{figure}[H]
    \centerline{\includegraphics[scale= 0.60]{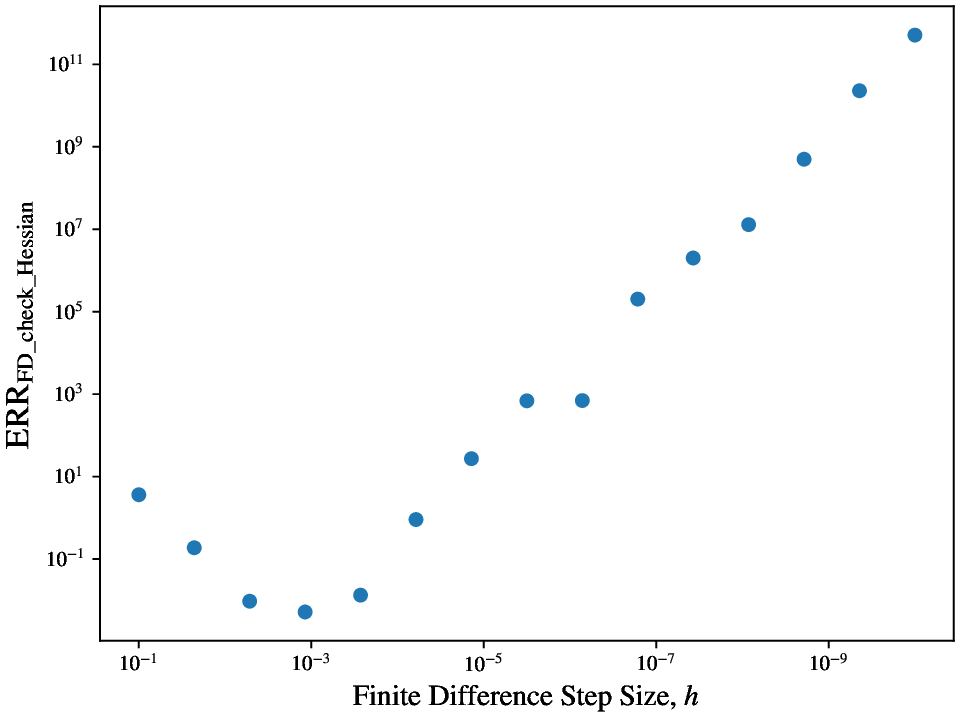}}
    \caption{Direct-adjoint Hessian comparisons to finite difference Hessian}
    \label{fig:FD_Hessian}
\end{figure}

First, noiseless synthetic data are considered to check the correctness of the gradients and Hessians computed using the direct and adjoint approaches. This is accomplished by choosing a direction vector~$\bd$ and computing the absolute error between the directional derivatives of the local sensitivity analysis (LSA) gradients and Hessians, and those found using finite difference schemes. In this study, each component of the direction vector is randomly chosen such that~$d_i,\ i = 1, 2, 3$ (corresponding to the three active design variables) is drawn from a uniform distribution $U[-1,1]$. Concretely, 
the error for gradients is expressed as 
\begin{equation}
    \text{ERR}_{\text{FD\_check\_grad}}\ =\ \left \lvert
\left(\left[\frac{dJ}{d\bp}\right]_{\text{LSA}} \cdot \bd \right) -
\left(\left[\frac{dJ}{d\bp}\right]_{\text{FD}} \cdot \bd \right)\right \rvert\
.
\end{equation}
The error for the objective Hessian is computed similarly and given by 
\begin{equation}
    \text{ERR}_{\text{FD\_check\_Hessian}}\ =\ \left \lvert \left(\bd^T
\left[\frac{d^2J}{d\bp^2}\right]_{\text{LSA}} \bd \right) - \left(\bd^T
\left[\frac{d^2J}{d\bp^2}\right]_{\text{FD}}  \bd \right)\right \rvert\ .
\end{equation}
The preceding errors are computed using a sequence of decreasing finite difference step sizes 
${h = 10^{-k}\ ,\ k = 2,\, \dots\, ,10}$. The results of the finite difference checks 
are plotted in Figures~\ref{fig:FD_grad} and~\ref{fig:FD_Hessian} for the gradient and Hessian, respectively.
\par
For both the gradient and the Hessian, it is observed that the error in the finite-difference approximation decreases monotonically up to an inflection point. After this point, 
floating-point cancellation errors begin to grow, causing the finite-difference 
approximation to lose accuracy. This well-known behavior provides strong evidence that the gradients and Hessians calculated through AD approaches are exact. 
To provide further verification, the complex-step derivative method, which circumvents the loss of accuracy due to floating-point cancellation error \cite{lai}, is also implemented. The results pertaining to the complex-step method are found in Appendix \ref{App:complex_step}.
\par
\begin{figure}[H] 
    \centerline{\includegraphics[scale= 0.40]{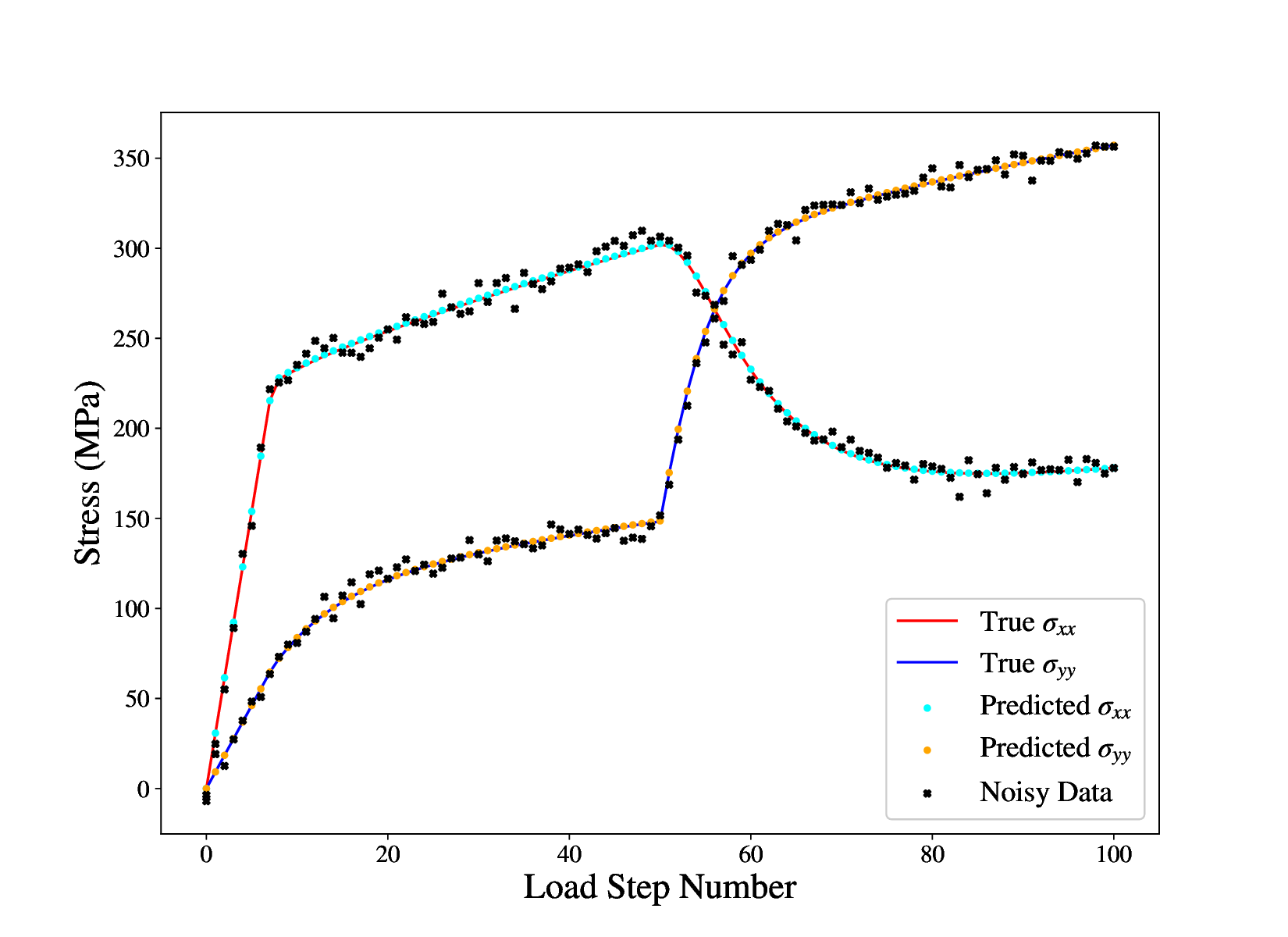}}
    \caption{Comparison of calibrated model with synthetic data for 
    $\sigma_{noise} = 5$ MPa}
    \label{fig:synth_stress_5}
\end{figure}
\begin{figure}[H] 
    \centerline{\includegraphics[scale= 0.40]{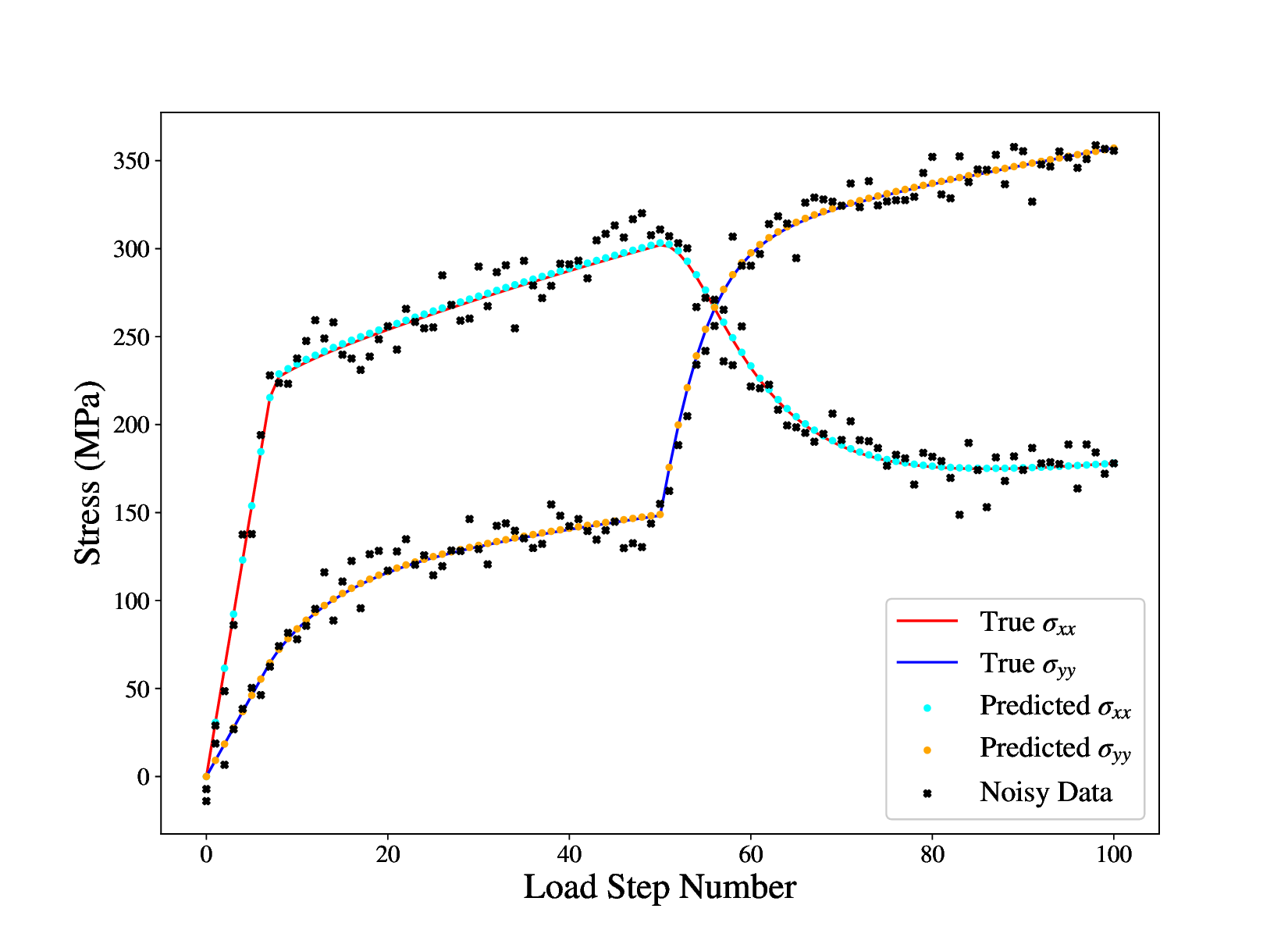}}
    \caption{Comparison of calibrated model with synthetic data for 
    $\sigma_{noise} = 10$ MPa}
    \label{fig:synth_stress_10}
\end{figure}
\begin{figure}[H]
    \centerline{\includegraphics[scale= 0.60]{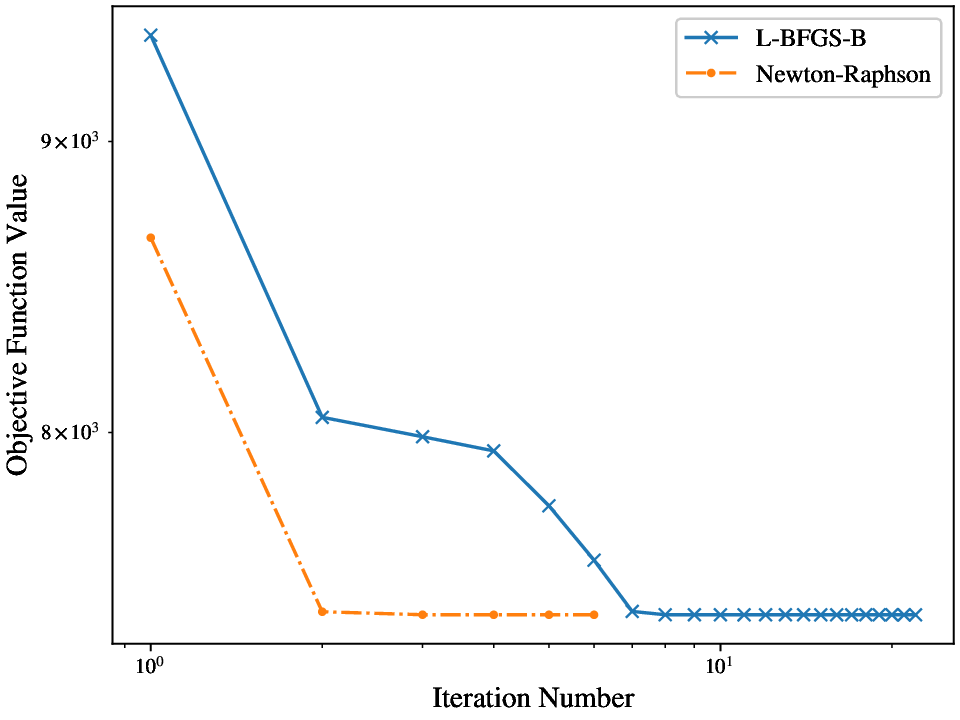}}
    \caption{Convergence of plane-stress optimization problem for 
    $\sigma_{noise} = 5$ MPa}
    \label{fig:conv_5}
\end{figure}
\begin{figure}[H]
    \centerline{\includegraphics[scale= 0.60]{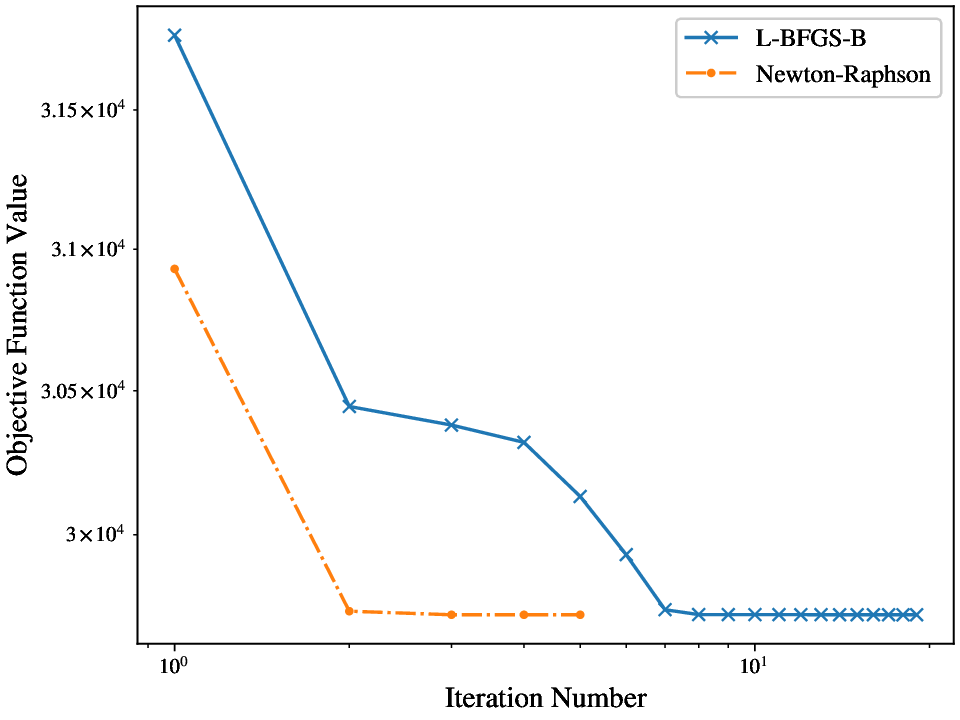}}
    \caption{Convergence of plane-stress optimization problem for 
    $\sigma_{noise} = 10$ MPa}
    \label{fig:conv_10}
\end{figure}

For another test, an initial guess $\bp_0$ of the material parameters  slightly different from the true material 
parameters is set to $[E, \nu, Y, K, S, D] = [70000, \, 0.3, \, 220, \, 0, \, 220, \, 22]$. Again, $E$, $Y$, $K$, and $S$ are written in units of MPa.
Then, the L-BFGS-B and Newton-Raphson algorithms are used to determine the optimal material parameters that minimize the objective function in Equation~\eqref{eq:opt_eqn}. Both algorithms 
are set to terminate when $\left\Vert\frac{dJ}{d\bp}\right\Vert_{\infty} < 10^{-4}$. Up to floating-point precision, there was no difference between the model parameters used to generate the synthetic data and those obtained from the solution of the optimization problem.
\begin{table}[H] 
\centering 
\rowcolors{2}{gray!25}{white}
\scriptsize
\hspace{-0.2in}\begin{tabular}{c c c c c c c}
\rowcolor{gray!75}
 & $Y$ & $S$ & $D$ & Iterations & \shortstack{Time \\
 sec} &\shortstack{Total Number of \\ System Solutions}\\
Truth & 200 & 200 & 20 & NA & NA & NA\\ Initial & 220 & 220 & 22 & NA & NA & NA\\ 
\multicolumn{7}{c}{\begin{tabular}{l}
$\sigma_{\text {noise }}\ =\ 5$
\end{tabular}} \\
L-BFGS-B & 200.781 (0.391\%) & 195.638 (2.181\%) & 20.4537 (2.268\%) & 22 & 4.6 & 10520\\ 
Newton-Raphson & 200.781 (0.391\%) & 195.638 (2.181\%) & 20.4537 (2.268\%) & 6 & 2.2 & 5162 
\\ 
\multicolumn{7}{c}{\begin{tabular}{l}
$\sigma_{\text {noise }}\ =\ 10$
\end{tabular}} \\
L-BFGS-B & 201.5584 (0.779\%) & 191.425 (4.288\%) & 20.9129 (4.565\%) & 19 & 3.8 &9605\\ 
Newton-Raphson & 201.5584 (0.779\%) & 191.425 (4.288\%) & 20.9129 (4.565\%) & 5 & 1.9 & 4302
\\ 
\end{tabular}
\begin{tablenotes}
      \small
      \item \textit{Note}: Errors relative to the true solutions are displayed next to the parameter values
    \end{tablenotes}
\caption {Inverse problem solutions for plane stress problem with noisy synthetic data} \label{tab:inverse_sol_synthetic}
\end{table}
Finally, two sets of noisy stress-strain data are considered, each created 
by adding normally distributed random variables with zero mean and standard deviations 
$\sigma_{\text{noise}} = 5$ and $\sigma_{\text{noise}} = 10$~MPa to the stress 
of each synthetic data point. The optimization problem is solved for both noisy 
datasets using L-BFGS-B and Newton-Raphson algorithms. The algorithms are again set to terminate when $\left\Vert\frac{dJ}{d\bp}\right\Vert_{\infty} < 10^{-4}$. 
The results obtained here are summarized in Table~\ref{tab:inverse_sol_synthetic}. Note from the table that the errors relative to the true solutions are directly proportional to the amount of noise added to the data. To visualize the calibration accuracy, the optimized solution is overlaid on the true solution in Figures~\ref{fig:synth_stress_5} and~\ref{fig:synth_stress_10}. 
Finally, convergence plots are shown in Figure~\ref{fig:conv_5} and~\ref{fig:conv_10}. It is observed that both methods arrive at 
\noindent  
the same optimal solution; however, the Newton-Raphson algorithm 
requires approximately one-third the number of iterations compared to the L-BFGS-B algorithm. More information on the convergence of the Newton-Raphson algorithm for this numerical example is provided in Table~\ref{tab:data_synthetic}. The computational cost is determined based on the total number of system solutions, including those required to solve the forward problem, and those needed to compute the gradients and Hessians. Although each iteration of the Newton-Raphson algorithm incurs a higher cost, the L-BFGS-B algorithm requires nearly twice the number of total system solutions due to the high number of iterations.         
\subsection{Calibration with experimental tension data}
\par\noindent
After verifying the model calibration process with synthetic data, the calibration 
is performed on actual experimental data. The aim is to obtain the optimal plasticity parameters of 304L stainless steel by utilizing stress-strain data from a uniaxial tensile test \cite{jones_digital_2024}. This is conducted on a tensile dog-bone specimen 
with a gauge section of 50.8 mm in length, 12.7 mm in width, and 1.5 mm in thickness. 
During the test, a load cell records the tensile force and digital image correlation
(DIC) is used to measure axial strain using a virtual extensometer in the gauge
section of the specimen. Since the model presented in this study is applicable only to infinitesimal deformations, force measurements at points greater than 0.02 mm/mm strain are 
omitted in the calibration process.
\par
Under the assumption that the stress is uniaxial
(see Appendix~\ref{App:uniaxial} for enforcement of this condition), the optimization problem
Equation~\eqref{eq:opt_eqn} is solved using the L-BFGS-B and Newton-Raphson 
algorithms. The gradient and Hessian at each load step are computed using the 
adjoint and direct-adjoint approaches, respectively. For these model calibration runs, the 
elastic parameters~$E$ and~$\nu$ are assumed to be known and fixed to values of $183$~GPa 
and $0.29$, respectively. Again, both optimization algorithms are set to 
terminate when $\left\Vert\frac{dJ}{d\bp}\right\Vert_{\infty} < 10^{-4}$. The initial material parameters for 
the inverse problem are determined using a genetic algorithm (GA) 
\cite{10.5555/534133}. The GA starts with a population of 25 randomly generated sets of material parameters $\bp$, which in the context of the algorithm are referred to as \textit{chromosomes}. The cost for each chromosome is determined by evaluating the objective function $J$. The 8 chromosomes with the lowest cost are selected to become the \textit{parents}. Following the method outlined in \cite{10.5555/534133}, pairs of parent chromosomes are \textit{bred} to create 8 \textit{child} chromosomes. Each new \textit{generation} consists of the 8 parents from the previous generation, their offspring, and 9 newly generated chromosomes. The GA is run for 4 generations, which equates to a total of 100 objective function evaluations. The initial guess for the parameters is ultimately chosen to be $[Y, \, K, \, S, \, D] = 
[140, \, 3400, \, 180, \, 1500]$. Here, $Y$, $K$, and $S$ are again written in units of MPa.  
\begin{table}[H]
\centering 
\rowcolors{2}{gray!25}{white}
\scriptsize
\begin{tabular}{c c c c c c c c}
\rowcolor{gray!75}
 & $Y$ &$K$ & $S$ & $D$ & Iterations & \shortstack{Time \\
 sec} &\shortstack{Total Number of \\ System Solutions}\\
Initial & 140 & 3400 & 180 & 2500 & NA & NA & NA\\ L-BFGS-B & 148.223 & 3472.982 & 
177.961 & 2589.541 & 20 & 35 & 39467 \\ Newton-Raphson & 148.223 & 3472.982 & 177.961 & 
2589.541 & 8 & 29 & 33209\\ 
\end{tabular}
\caption {Inverse Problem solutions for uniaxial stress problem with experimental data} \label{tab:inverse_sol} 
\end{table}
\par
\begin{figure}[H]
    \centerline{\includegraphics[scale=0.70]{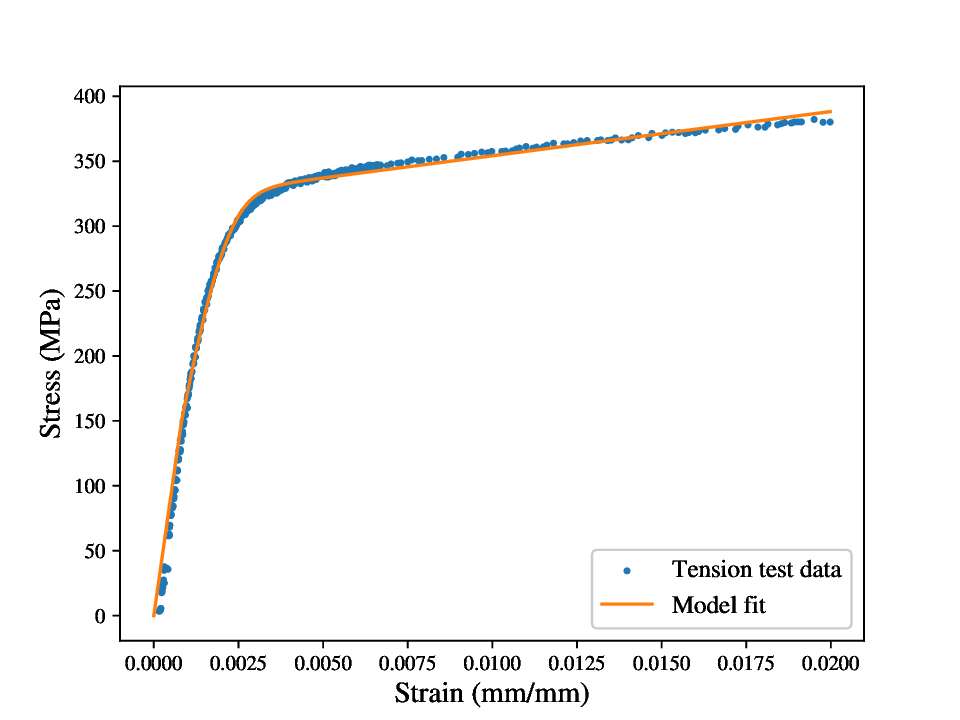}}
    \caption{Comparison of calibrated model with experimental data}
    \label{fig:exp_calibration}
\end{figure}

The calibration results are summarized in Table~\ref{tab:inverse_sol}, and an overlay 
of the optimized model stress-strain on the experimental data is plotted 
in Figure~\ref{fig:exp_calibration}. Additional information pertaining to convergence is presented in Table~\ref{tab:data uniaxial}. Similar to results from the synthetic 
example, the optimal parameters determined by both algorithms are identical. 
The L-BFGS-B algorithm however, requires more than double the number of 
iterations compared to the Newton-Raphson algorithm. 
%
\section{Conclusions}\label{sec:con}
\par\noindent
 This article proposes a direct-adjoint approach to compute the Hessian for the calibration of elastoplastic material parameters formulated as a constrained optimization problem. A key step in this approach is the derivation of the second-order derivatives required for the Hessian calculation. By leveraging AD, these derivatives can be computed efficiently, which eliminates the need for tedious analytical derivations.
\par 
The calibration process is first tested on a synthetic plane-stress problem to confirm the accuracy of the derivations and implementation. Indeed, the Hessian matrix obtained through the direct-adjoint approach compares as desired to the one computed using the finite difference method. A subsequent calibration is performed on experimental data from a uniaxial tension test. In both numerical examples, the Newton-Raphson algorithm consistently demonstrated a faster convergence rate compared to quasi-Newton gradient-based algorithms such as L-BFGS-B, resulting in a reduced overall computational cost. The findings of this work establish a solid foundation for the future development of sensitivity-based methods for material parameter calibration from full-scale (as opposed to simply material point) calibration problems with FE model constraints.

\protect\vspace{0.1in}
\begin{center}
{\large\bf Acknowledgments}
\end{center}
\nopagebreak[5]
This work was supported by the Laboratory Directed Research and Development program at Sandia National Laboratories, a multimission laboratory managed and operated by National Technology and Engineering Solutions of Sandia LLC (NTESS), a wholly owned subsidiary of Honeywell International Inc. for the U.S.  Department of Energy’s National Nuclear Security Administration under contract DE-NA0003525.  This written work is authored by an employee of NTESS. The employee, not NTESS, owns the right, title and interest in and to the written work and is responsible for its contents. Any subjective views or opinions that might be expressed in the written work do not necessarily represent the views of the U.S. Government. The publisher acknowledges that the U.S. Government retains a non-exclusive, paid-up, irrevocable, world-wide license to publish or reproduce the published form of this written work or allow others to do so, for U.S. Government purposes. The DOE will provide public access to results of federally sponsored research in accordance with the DOE Public Access Plan.
\section*{Data Availability}
\par\noindent
Data will be made available on request.
\newpage 
\ssp
\addcontentsline{toc}{section}{\string\numberline{}References}
\bibliography{refs}

\begin{thebibliography}{10}

\bibitem{ma16020836}
Y.~Hou, D.~Myung, J.~K. Park, J.~Min, H.-R. Lee, A.~A. El-Aty, and M.-G. Lee,
  ``A review of characterization and modelling approaches for sheet metal
  forming of lightweight metallic materials,'' {\em Materials}, vol.~16, no.~2,
  2023.

\bibitem{article}
S.~Avril, M.~Bonnet, C.-B. A.S, M.~Grédiac, F.~Hild, P.~Ienny, F.~Latourte,
  D.~Lemosse, S.~Pagano, E.~Pagnacco, and F.~Pierron, ``Overview of
  identification methods of mechanical parameters based on full-field
  measurements,'' {\em Experimental Mechanics}, vol.~48, 08 2008.

\bibitem{ROBSON2024104881}
J.~D. Robson, D.~Armstrong, J.~Cordell, D.~Pope, and T.~F. Flint, ``Calibration
  of constitutive models using genetic algorithms,'' {\em Mechanics of
  Materials}, vol.~189, p.~104881, 2024.

\bibitem{PAL1996325}
S.~Pal, G.~{Wije Wathugala}, and S.~Kundu, ``Calibration of a constitutive
  model using genetic algorithms,'' {\em Computers and Geotechnics}, vol.~19,
  no.~4, pp.~325--348, 1996.

\bibitem{ROKONUZZAMAN2010573}
M.~Rokonuzzaman and T.~Sakai, ``Calibration of the parameters for a
  hardening–softening constitutive model using genetic algorithms,'' {\em
  Computers and Geotechnics}, vol.~37, no.~4, pp.~573--579, 2010.

\bibitem{Levasseur}
S.~Levasseur, Y.~Malécot, M.~Boulon, and E.~Flavigny, ``Soil parameter
  identification using a genetic algorithm,'' {\em International Journal for
  Numerical and Analytical Methods in Geomechanics}, vol.~32, no.~2,
  pp.~189--213, 2008.

\bibitem{papon}
A.~Papon, Y.~Riou, C.~Dano, and P.-Y. Hicher, ``Single-and multi-objective
  genetic algorithm optimization for identifying soil parameters,'' {\em
  International Journal for Numerical and Analytical Methods in Geomechanics},
  vol.~36, no.~5, pp.~597--618, 2012.

\bibitem{doi:10.1061/(ASCE)EM.1943-7889.0001214}
C.~Smith, A.~Kanvinde, and G.~Deierlein, ``Calibration of continuum cyclic
  constitutive models for structural steel using particle swarm optimization,''
  {\em Journal of Engineering Mechanics}, vol.~143, no.~5, p.~04017012, 2017.

\bibitem{https://doi.org/10.1002/nag.137}
H.~Mattsson, M.~Klisinski, and K.~Axelsson, ``Optimization routine for
  identification of model parameters in soil plasticity,'' {\em International
  Journal for Numerical and Analytical Methods in Geomechanics}, vol.~25,
  no.~5, pp.~435--472, 2001.

\bibitem{CEKEREVAC2006432}
C.~Cekerevac, S.~Girardin, G.~Klubertanz, and L.~Laloui, ``Calibration of an
  elasto-plastic constitutive model by a constrained optimisation procedure,''
  {\em Computers and Geotechnics}, vol.~33, no.~8, pp.~432--443, 2006.

\bibitem{app9091799}
X.~Zhao, Y.~Sun, and Y.~Mei, ``A size-dependent cost function to solve the
  inverse elasticity problem,'' {\em Applied Sciences}, vol.~9, no.~9, 2019.

\bibitem{SAKARIDIS2024113076}
E.~Sakaridis, C.~C. Roth, B.~Jordan, and D.~Mohr, ``Post-necking full-field
  femu identification of anisotropic plasticity from flat notched tension
  experiments,'' {\em International Journal of Solids and Structures},
  vol.~305, p.~113076, 2024.

\bibitem{Yueqi}
Y.~Wang, S.~Coppieters, P.~Lava, and D.~Debruyne, ``Anisotropic yield surface
  identification of sheet metal through stereo finite element model updating,''
  {\em The Journal of Strain Analysis for Engineering Design}, vol.~51, no.~8,
  pp.~598--611, 2016.

\bibitem{VARGAS20226733}
R.~Vargas, R.~Canto, and F.~Hild, ``Cohesive properties of refractory castable
  at 600°c: Effect of sintering and testing temperature,'' {\em Journal of the
  European Ceramic Society}, vol.~42, no.~14, pp.~6733--6749, 2022.

\bibitem{Cooreman}
S.~Cooreman, D.~Lecompte, H.~Sol, J.~Vantomme, and D.~Debruyne,
  ``Identification of mechanical material behavior through inverse modeling and
  {DIC},'' {\em Experimental Mechanics}, vol.~48, pp.~421--433, 08 2008.

\bibitem{seidl_calibration_2022}
D.~T. Seidl and B.~N. Granzow, ``Calibration of elastoplastic constitutive
  model parameters from full-field data with automatic differentiation-based
  sensitivities,'' {\em International Journal for Numerical Methods in
  Engineering}, vol.~123, no.~1, pp.~69--100, 2022.
\newblock eprint: https://onlinelibrary.wiley.com/doi/pdf/10.1002/nme.6843.

\bibitem{numericalopt}
J.~Nocedal and S.~Wright, {\em Numerical Optimization}.
\newblock Springer Series in Operations Research and Financial Engineering,
  Springer, 2~ed., 2006.

\bibitem{NEGGERS201971}
J.~Neggers, F.~Mathieu, F.~Hild, and S.~Roux, ``Simultaneous full-field
  multi-experiment identification,'' {\em Mechanics of Materials}, vol.~133,
  pp.~71--84, 2019.

\bibitem{ARIAN1999853}
E.~Arian and S.~Ta’asan, ``Analysis of the hessian for aerodynamic
  optimization: inviscid flow,'' {\em Computers \& Fluids}, vol.~28, no.~7,
  pp.~853--877, 1999.

\bibitem{SHIDONG2018265}
D.~Shi-Dong and S.~Nadarajah, ``Approximate hessian for accelerated convergence
  of aerodynamic shape optimization problems in an adjoint-based framework,''
  {\em Computers \& Fluids}, vol.~168, pp.~265--284, 2018.

\bibitem{lame_manual}
B.~T. Lester, M.~R. Buche, K.~N. Long, W.~M. Scherzinger, K.~N. Cundiff,
  B.~Reedlunn, C.~Hamel, and A.~J. Stershic, ``Library of advanced materials
  for engineering (lamÉ) 5.22,'' tech. rep., Sandia National Lab. (SNL-NM),
  Albuquerque, NM (United States), 10 2024.

\bibitem{alberdi_unified_2018}
R.~Alberdi, G.~Zhang, L.~Li, and K.~Khandelwal, ``A unified framework for
  nonlinear path-dependent sensitivity analysis in topology optimization,''
  {\em International Journal for Numerical Methods in Engineering}, vol.~115,
  no.~1, pp.~1--56, 2018.
\newblock \_eprint: https://onlinelibrary.wiley.com/doi/pdf/10.1002/nme.5794.

\bibitem{papadimitriou}
D.~I. Papadimitriou and K.~C. Giannakoglou, ``Direct, adjoint and mixed
  approaches for the computation of hessian in airfoil design problems,'' {\em
  International Journal for Numerical Methods in Fluids}, vol.~56, no.~10,
  pp.~1929--1943, 2008.

\bibitem{computationalplasticity}
E.~A. de~Souza~Neto, D.~Peric, and D.~R. Owen, {\em Computational methods for
  plasticity: theory and applications}.
\newblock John Wiley \& Sons, 2011.

\bibitem{SimoHughes}
J.~Simo and T.~Hughes, {\em Computational Inelasticity}.
\newblock Interdisciplinary Applied Mathematics, Springer New York, 2006.

\bibitem{jax}
J.~Bradbury, R.~Frostig, P.~Hawkins, M.~J. Johnson, C.~Leary, D.~Maclaurin,
  G.~Necula, A.~Paszke, J.~Vander{P}las, S.~Wanderman-{M}ilne, and Q.~Zhang,
  ``{JAX}: composable transformations of {P}ython+{N}um{P}y programs,'' 2024.

\bibitem{scipy}
P.~Virtanen, R.~Gommers, T.~E. Oliphant, M.~Haberland, T.~Reddy, D.~Cournapeau,
  E.~Burovski, P.~Peterson, W.~Weckesser, J.~Bright, S.~J. {van der Walt},
  M.~Brett, J.~Wilson, K.~J. Millman, N.~Mayorov, A.~R.~J. Nelson, E.~Jones,
  R.~Kern, E.~Larson, C.~J. Carey, {\.I}.~Polat, Y.~Feng, E.~W. Moore,
  J.~{VanderPlas}, D.~Laxalde, J.~Perktold, R.~Cimrman, I.~Henriksen, E.~A.
  Quintero, C.~R. Harris, A.~M. Archibald, A.~H. Ribeiro, F.~Pedregosa, P.~{van
  Mulbregt}, and {SciPy 1.0 Contributors}, ``{{SciPy} 1.0: Fundamental
  Algorithms for Scientific Computing in Python},'' {\em Nature Methods},
  vol.~17, pp.~261--272, 2020.

\bibitem{lai}
K.-L. Lai and J.~Crassidis, ``Extensions of the first and second complex-step
  derivative approximations,'' {\em Journal of Computational and Applied
  Mathematics}, vol.~219, no.~1, pp.~276--293, 2008.

\bibitem{jones_digital_2024}
E.~M.~C. Jones, P.~L. Reu, S.~L.~B. Kramer, A.~R. Jones, J.~D. Carroll, K.~N.
  Karlson, D.~T. Seidl, and D.~Z. Turner, ``Digital image correlation and
  infrared thermography data for seven unique geometries of {304L} stainless
  steel,'' {\em Scientific Data}, vol.~11, p.~1101, Oct. 2024.
\newblock Publisher: Nature Publishing Group.

\bibitem{10.5555/534133}
D.~E. Goldberg, {\em Genetic Algorithms in Search, Optimization and Machine
  Learning}.
\newblock USA: Addison-Wesley Longman Publishing Co., Inc., 1st~ed., 1989.

\bibitem{lyness1967}
J.~N. Lyness and C.~B. Moler, ``Numerical differentiation of analytic
  functions,'' {\em SIAM Journal on Numerical Analysis}, vol.~4, no.~2,
  pp.~202--210, 1967.

\bibitem{lyness1968}
J.~Lyness, ``Differentiation formulas for analytic functions,'' {\em
  Mathematics of Computation}, vol.~22, no.~102, pp.~352--362, 1968.

\end{thebibliography}
\dsp
%
%
\setcounter{section}{0}
\setcounter{equation}{0}
\renewcommand{\thesection}{\Alph{section}}
\refstepcounter{section}
\def\theequation{\Alph{section}.\arabic{equation}}
\addcontentsline{toc}{section}{Appendix A: Plane Stress and Uniaxial Stress Models From 3D}
\section*{Appendix A: Plane Stress and Uniaxial Stress Models From 3D}\label{App:plane_uniaxial}
\par\noindent
\subsection{Plane stress} \label{App:plane}
\par\noindent
A material point is said to be in plane stress if the only non-zero stress 
components act in one plane only. Thus, for a plane with unit normal~$\be_3$, 
where vectors \{$\be_1$,~$\be_2$,~$\be_3$\} form an orthonormal basis, it is 
assumed that $\sigma_{13} = \sigma_{23} = \sigma_{33} = 0$. For isotropic 
linearly elastic materials, the condition, $\sigma_{13} = \sigma_{23} = 0$, 
is automatically satisfied if $\epsilon^e_{13} = \epsilon^e_{23} = 0$. 
However, to enforce~$\sigma_{33} = 0$, it follows from 
Equation~\eqref{eq:elastic_stress} that there may exist a non-zero off-axis strain, 
$\epsilon^e_{33}$. Implementing the plane stress constraints in the model 
requires two modifications. First, an off-axis stretch in the~$\be_3$-direction 
is added to the local state variables $\Bxi^n_{\text{plane}}$, which now takes the form 
\begin{equation}
    [\epsilon^p_{11},\ \epsilon^p_{12},\
\epsilon^p_{13}, \epsilon^p_{22},\ \epsilon^p_{23},\ \epsilon^p_{33},\ \alpha,\
\epsilon_{33}]^{n}, \quad n\ =\ 1,\dots,N_L\ .
\end{equation}
Second, an extra residual equation is added to enforce the condition 
$\sigma_{33} = 0$, such that 
\begin{equation}
\bC^n_{\text{plane}}\ = \ \left\{
\begin{aligned}
&f\ <\ 0: \left\{
\begin{aligned}
  &\left(\Bepsilon^p\right)^n - \left(\Bepsilon^p\right)^{n-1}\ , \\
  &\alpha^n - \alpha^{n-1}\\
  &\lambda \, \text{tr} \left[\Bepsilon \right] + 2 \mu \left(\epsilon_{33} -\epsilon^p_{33}\right)\ ,
\end{aligned} \right. \\
&f\ =\ 0: \left\{
\begin{aligned}
  &\left(\Bepsilon^p\right)^n - \left(\Bepsilon^p\right)^{n-1}
  - \left(\alpha^n - \alpha^{n-1}\right)\bn^n\ , \\
  &f(\Bsigma^n, \alpha^n)\\
  &\lambda \, \text{tr} \left[\Bepsilon \right] + 2 \mu \left(\epsilon_{33} -\epsilon^p_{33}\right)\ .
\end{aligned} \right.
\end{aligned}
\right.
\end{equation}
\subsection{Uniaxial stress} \label{App:uniaxial}
\par\noindent
For the case 
of simple tension or compression along the~$\be_1$-axis, $\sigma_{11} \neq 0$, while all other components of stress are zero. In an isotropic linearly 
elastic solid, stress components $\sigma_{12}$, $\sigma_{13}$, and 
$\sigma_{23}$ automatically vanish if $\epsilon^e_{12} = \epsilon^e_{13} = 
\epsilon^e_{23} = 0$. To satisfy the constraint $\sigma_{22} = \sigma_{33}
= 0$, it follows from inverting Equation~\eqref{eq:elastic_stress} that 
\begin{equation}
    \epsilon^e_{22}\ =\ \epsilon^e_{33}\ = -\frac{\nu \sigma_{11}}{E}\ .
\end{equation}
Similar to the plane stress case, off-axis stretches parallel to~$\be_2$ and 
$\be_3$ are added to the local state variables $\Bxi^n_{\text{uniaxial}}$, which one may interpret as 
\begin{equation}
    [\epsilon^p_{11},\ \epsilon^p_{12},\ \epsilon^p_{13}, \epsilon^p_{22},\ \epsilon^p_{23},\ \epsilon^p_{33},\ \alpha,\ \epsilon_{22},\ \epsilon_{33}]^{n} , \quad n\ =\ 1,\dots,N_L\ .
\end{equation}
To determine the correct values of~$\epsilon_{22}$ and~$\epsilon_{33}$ at each load step, 
constraint equations $\sigma_{22} = \sigma_{33} = 0$ are added to the local 
residual equations. 
\begin{equation}
\bC^n_{\text{uniaxial}}\ = \ \left\{
\begin{aligned}
&f\ <\ 0: \left\{
\begin{aligned}
  &\left(\Bepsilon^p\right)^n - \left(\Bepsilon^p\right)^{n-1}\ , \\
  &\alpha^n - \alpha^{n-1}\ ,\\
  &\lambda \, \text{tr} \left[\Bepsilon \right] + 2 \mu \left(\epsilon_{\beta\beta} -\epsilon^p_{\beta\beta}\right)\ , \quad \beta\ =\ 2,\ 3\ ,
\end{aligned} \right. \\
&f\ =\ 0: \left\{
\begin{aligned}
  &\left(\Bepsilon^p\right)^n - \left(\Bepsilon^p\right)^{n-1}
  - \left(\alpha^n - \alpha^{n-1}\right)\bn^n\ , \\
  &f(\Bsigma^n, \alpha^n)\ ,\\
  &\lambda \, \text{tr} \left[\Bepsilon \right] + 2 \mu \left(\epsilon_{\beta\beta} -\epsilon^p_{\beta\beta}\right)\ , \quad \beta\ =\ 2,\ 3\ .
\end{aligned} \right.
\end{aligned}
\right.
\end{equation}
\addcontentsline{toc}{section}{Appendix B: Optimization Details for the Numerical Examples}
\setcounter{equation}{0}
\setcounter{table}{0}
\refstepcounter{section}
\def\theequation{\Alph{section}.\arabic{equation}}
\renewcommand\thetable{\Alph{section}\arabic{table}}
\section*{Appendix B: Optimization details for Numerical Examples}\label{App:details}
\par\noindent
This appendix contains tables of objective function histories and Hessian matrix condition numbers for each of the numerical examples presented in Section \ref{sec:numerical_results}.
\begin{table}[H] 
\centering 
\renewcommand{\arraystretch}{1.2}
\rowcolors{2}{gray!25}{white}
\scriptsize
\begin{tabular}{c c c c}
\rowcolor{gray!75}
 Iteration Number & $\log_{10}(J)$ & $\left\Vert\frac{dJ}{d\bp}\right\Vert$ &  cond$\left(\left[\frac{d^2J}{d\bp^2}\right]\right)$\\ 
\multicolumn{4}{c}{\begin{tabular}{l}
$\sigma_{\text {noise }}\ =\ 5$
\end{tabular}} \\
0 & 4.90065294 & $9.83 \times 10^{5}$ & $1.78 \times 10^{2}$ \\ 
1 & 3.93733757 & $1.12 \times 10^{5}$ & $1.30 \times 10^{3}$ \\
2 & 3.87159559 & $2.02 \times 10^{3}$ & $9.68 \times 10^{3}$ \\
3 & 3.87107456 & $9.88 \times 10^{1}$ & $7.14 \times 10^{3}$ \\
4 & 3.87106385 & $4.49 \times 10^{1}$ & $6.8 \times 10^{3}$ \\
5 & 3.87106383 & $1.30 \times 10^{-2}$ &  $6.83 \times 10^{3}$ \\
6 & 3.87106383 & $1.54 \times 10^{-7}$ &  $6.83 \times 10^{3}$
\\
\multicolumn{4}{c}{\begin{tabular}{l}
$\sigma_{\text {noise }}\ =\ 10$
\end{tabular}} \\
0 & 4.99960149 & $9.68 \times 10^{5}$ & $1.79 \times 10^{2}$ \\ 
1 & 4.49038115 & $1.09 \times 10^{5}$ & $1.28 \times 10^{3}$ \\
2 & 4.47330414 & $2.34 \times 10^{3}$ & $1.20 \times 10^{4}$ \\
3 & 4.47312501 & $2.19 \times 10^{2}$ & $6.69 \times 10^{3}$ \\
4 & 4.47312484 & $1.24 \times 10^{-1}$ & $6.57 \times 10^{3}$ \\
5 & 4.47312484 & $2.04 \times 10^{-6}$ &  $6.57 \times 10^{3}$
\\  
\end{tabular}
\caption {Newton-Raphson convergence data for synthetic plane stress example} \label{tab:data_synthetic}

\end{table}
\begin{table}[H] 
\centering 
\renewcommand{\arraystretch}{1.2}
\rowcolors{2}{gray!25}{white}
\scriptsize
\begin{tabular}{c c c c}
\rowcolor{gray!75}
 Iteration Number & $\log_{10}(J)$ & $\left\Vert\frac{dJ}{d\bp}\right\Vert$ &  cond$\left(\left[\frac{d^2J}{d\bp^2}\right]\right)$\\ 
0 & 4.44512889 & $3.82 \times 10^{5}$ & $9.63 \times 10^{2}$ \\ 
1 & 4.30036887 & $9.78 \times 10^{3}$ & $6.44 \times 10^{2}$ \\
2 & 4.29996754 & $1.20 \times 10^{2}$ & $7.11 \times 10^{2}$ \\
3 & 4.29996103 & $6.56 \times 10^{2}$ & $7.44 \times 10^{2}$ \\
4 & 4.29951592 & $8.73 \times 10^{2}$ & $7.89 \times 10^{2}$ \\
5 & 4.29912113 & $4.79 \times 10^{3}$ &  $6.47 \times 10^{2}$ \\
6 & 4.29908984 & $3.21 \times 10^{1}$ &  $6.98 \times 10^{2}$ \\
7 & 4.29908980 & $3.10 \times 10^{-1}$ &  $6.99 \times 10^{2}$ \\
8 & 4.29908980 & $5.56 \times 10^{-7}$ &  $6.99 \times 10^{2}$ \\
\end{tabular}
\caption {Newton-Raphson convergence data for uniaxial stress example}\label{tab:data uniaxial}
\end{table}
For both examples, the condition numbers tend to remain modest, which suggests that the underlying linear-algebraic problem is well-posed.
%
\addcontentsline{toc}{section}{Appendix C: Complex-step Derivative Checks}
\setcounter{equation}{0}
\refstepcounter{section}
\def\theequation{\Alph{section}.\arabic{equation}}
\section*{Appendix C: Complex-step Derivative Checks}\label{App:complex_step}
\par
\begin{figure}[H]
    \centerline{\includegraphics[scale= 0.85]{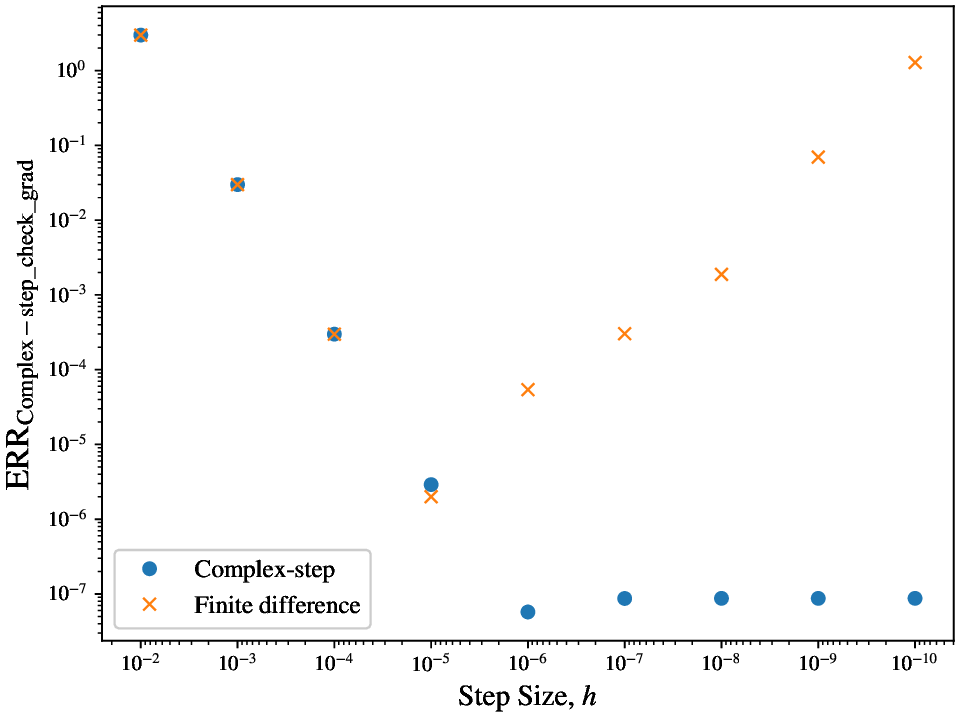}}
    \caption{Comparison between finite difference gradients and complex-step gradients}
    \label{fig:Complex_grad}
\end{figure}

\noindent
Complex-step derivative checks are implemented to further verify the 
correctness of the gradients and Hessian matrices computed using AD approaches. 
The first-order complex-step approximation for the first derivative is derived 
using a Taylor series expansion with a purely imaginary step size~$h$ 
\cite{lyness1967,lyness1968}. The scheme is written as 
\begin{equation}\label{eq:grad_complex_step}
    f'(x)\ =\ \frac{\Im[f(x + ih)]}{h} + O(h^2)\ .
\end{equation}
As observed from Equation~\eqref{eq:grad_complex_step}, only a single function 
evaluation is required, which avoids the floating-point cancellation errors 
typically seen in traditional finite difference methods. Results of the 
complex-step gradient check are shown in Figure~\ref{fig:Complex_grad}. Note 
that the absolute error of the gradient directional derivatives obtained using 
the complex step method reaches a minimum of~$10^{-7}$ at approximately~$h = 
10^{-6}$, and remains constant for subsequent smaller step-sizes. 
\par
\begin{figure}[h!]
    \centerline{\includegraphics[scale= 0.8]{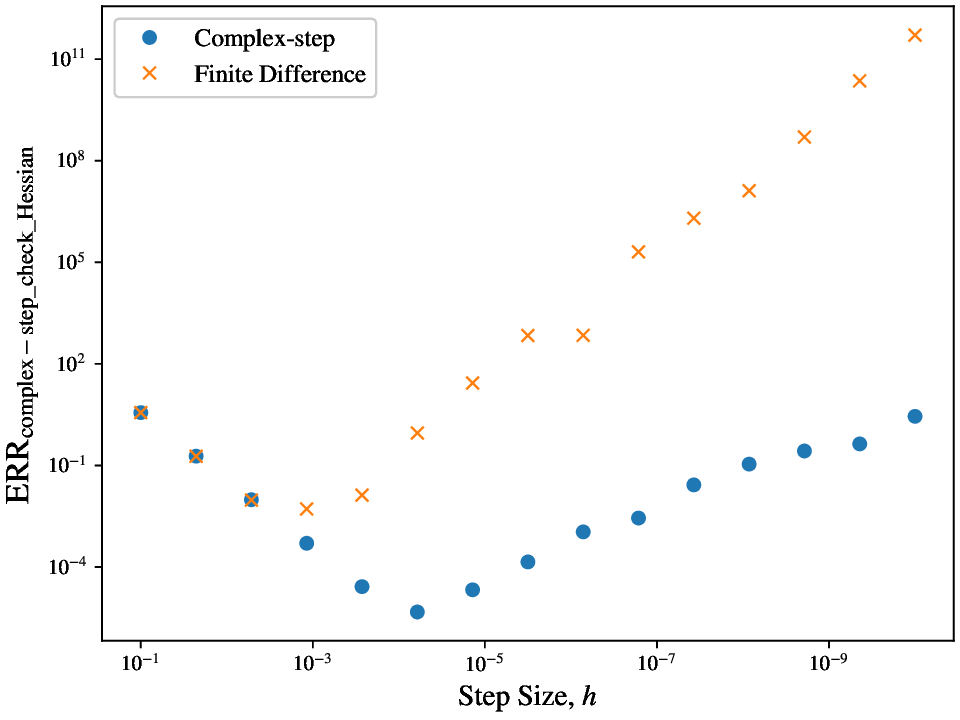}}
    \caption{Comparison between finite difference Hessian to complex-step Hessian} 
    \label{fig:Complex_Hessian}
\end{figure}

Derived using Richardson extrapolation, the second-order complex-step 
derivative scheme for obtaining the second derivative is expressed as 
\begin{equation}\label{eq:Hessian_complex_step}
    f''(x)\ =\ \frac{2\Im[f(x + i^{2/3}h) + f(x + i^{8/3}h)]}{\sqrt{3}h^2} + O(h^2)\ ,
\end{equation}
where $i^{a/b} = e^{i\theta}$ with phase angle~$\theta = 
\frac{p}{q}90^{\circ}$ \cite{lai}. The results for the complex-step Hessian 
check are presented in Figure~\ref{fig:Complex_Hessian}. One can observe that 
the errors in Hessian directional derivatives reaches a minimum of 
approximately~$10^{-5}$. For step-sizes smaller than~$h~=~10^{-5}$, the 
errors begin to increase due to round-off errors. This increase however occurs 
at a slower rate compared to the standard finite difference method. 
\par
\clearpage
\end{document}